# Light-Emitting Microfibers from Lotus Root for Eco-friendly Optical Waveguides and Biosensing

Xianguang Yang[1]*, Liping Xu[1], Shijie Xiong[1], Hao Rao[1], Fangchang Tan[1], Jiahao Yan[1], Yanjun Bao[1], Annachiara Albanese[2], Andrea Camposeo[3], Dario Pisignano[2,3], and Baojun Li[1]

[1]Guangdong Provincial Key Laboratory of Nanophotonic Manipulation, Institute of Nanophotonics, Jinan University, Guangzhou 511443, China

[2]Dipartimento di Fisica, Università di Pisa, Largo B. Pontecorvo 3, I-56127 Pisa, Italy

[3]NEST, Istituto Nanoscienze-CNR and Scuola Normale Superiore, Piazza S. Silvestro 12, I-56127 Pisa, Italy

**Abstract:** Optical biosensors based on micro-/nano-fibers are highly valuable for probing and monitoring liquid environments and bioactivity. Most of current optical biosensors, however, are still based on glass, semiconductors, or metallic materials, which might be not fully suited for biologically-relevant environments. Here, we introduce biocompatible and flexible microfibers from Lotus silk as micro-environmental monitors that exhibit waveguiding of intrinsic fluorescence as well as of coupled light. These features make single-filament monitors excellent building blocks for a variety of sensing functions, including pH-probing and detection of bacterial activity. These results pave the way for the development of new and entirely eco-friendly, potentially multiplexed biosensing platforms.

The continuous monitoring of the physio-pathological status of living organisms is highly important in modern biophysics and biomedicine, since even small fluctuations on top of baseline parameters might be signals of relevant biological or functional variations of cellular response. To address bio-environmental research and diagnostics towards on-body sensing, micro- and nano-optical biosensors have emerged as powerful tool [1-3]. These components have shown high sensitivity and rapid response due to their small size and large surface-to-volume ratio [4]. Micro/nano-optical waveguides and sensors based on them have been largely based on glass, metals and semiconductors [5-8]. For example, the photoelectric detection of spinal cord circuits was achieved by using Ag nanowire-coated polymer fiber probes [9], and silicon nanowire arrays were designed and used for the detection of C-reactive protein and hepatitis B virus [10, 11]. These systems show good performance, low transmission loss and high sensing efficiency, but they might impair cellular activity, which can strongly limit their biocompatibility and effective use in biomedical applications. Polymeric materials such as poly(methyl methacrylate) and polystyrene are also used to make sensors [12, 13, 14]. However, the fibers made of these synthetic polymers are nondegradable. Alternatively, biodegradable polymers can be used, such as polylactic acid [15]. However, the production cost for synthetic polymers can be high. Most of them do not feature intrinsic fluorescence, thus requiring chemical modification for achieving light emission, as well as numerous additional synthesis processes for obtaining fibers [16, 17], overall significantly limiting their applications in the biomedical field. In this respect, natural flexible materials with intrinsic fluorescence, able to guide light and enable optical sensing without interfering with cellular activity, are strongly desirable, especially if featuring very simple fabrication and processing [18, 19].

In this framework, natural fibers constitute a valuable choice for biosensing. Fibers can be extracted by various plants, such as the lotus root which is commonly used in daily life and it is very popular as a healthy food. It is composed of microfibers that are eco-friendly, flexible and wearable with excellent toughness [20, 21, 22], and well-suited for various applications in bioelectronics and medicine [23-25]. Some research reports mentioned that the natural lotus root filaments can serve as carriers for composite materials. The hydroxy and/or amino groups of natural lotus root filaments can interact with metal ions (e.g., $Mn^{2+}$, $Fe^{3+}$, $Co^{2+}$, $Ni^{2+}$, and $Cu^{2+}$) through electrostatic interactions. For example, metal-organic frameworks such as zeolitic imidazolate framework-67 were uniformly loaded onto the surface of natural lotus root filaments by simply mixing the precursors to prepare highly stable and active flexible electrocatalysts [26]. $Bi_2WO_6$ was chosen to prepare bismuth tungstate/natural lotus root filaments, for composite membranes with photocatalytic activity [27]. Moreover, passing white, green, and red light through the fibers of lotus stem revealed the ability of these natural fibers to transmit light [28]. Overall, there has been limited work on the investigation of the emission properties and light transport capabilities of single natural lotus root filaments, whereas their exploitation in biosensing remains largely unexplored.

Here we introduce lotus single-microfiber monitors of biological environments. The fibrous materials are extracted from the lotus root and exhibit cylindrical filaments with smooth surface and uniform diameter, which is highly beneficial to light guiding. They can emit broadband fluorescence, whose spectral properties can be tailored by varying the excitation wavelength. In addition, they transport the emitted light, as well as light from external sources coupled in them, over hundreds of micrometers with low loss. Based on waveguiding, we evidence sensing capacity for pH detection of fluids and monitoring bacterial activity. In particular, the sensitivity to local perturbations of the electromagnetic field evanescently coupled from the fiber to the environment is exploited for real-time monitoring of bacterial apoptotic process, through simple measurements of the varying transmitted light. These results lay the foundation for the exploitation of these natural single-filament microfibers in various, versatile and potentially multiplexed biosensing architectures.

Figure 1a shows a photograph of natural silks directly extracted by hands from lotus root. The length of silks is about 300 mm, which is three times as long as the original length. Figures 1b-c show scanning electron microscopy (SEM) images of micro-silks, extracted manually from natural Lotus roots and showing a well-defined, multi-filament helical structure [22]. Eight to thirteen finer single filaments with an average diameter around 4.2 μm are closely arranged in micro-silks (Figure 1d). Due to their uniform diameter and surface smoothness (Figure S1), these filaments can be good waveguides. Figures 1e-g show optical micrographs collected with long-pass filters upon laser excitation at 532, 440 and 365 nm, respectively. The isolated single-filaments are robust and flexible, and they can be micro-manipulated (see Experimental Section) to form complex geometries and loops (Figure S2). Figure 1h-j show optical micrographs of a loop-shaped single-filament microfiber, in which light from a 655 (h), 532 (i), or 440 (j) nm laser, respectively, is evanescently coupled by means of a tapered fiber and transmitted along the loop. In fact, the distance between the two ends of the microfiber is about 1 μm, which makes such architecture potentially exploitable as a resonant micro-cavity (see also Figure S3) [29]. We also performed further experiments on the fiber emission efficiency ($\eta$) emitted by this loop structure (Figure S4). The $\eta$ was calculated to be 18%, 20%, and 14% for red-, green-, and blue-emissive single-filament microfiber, respectively.

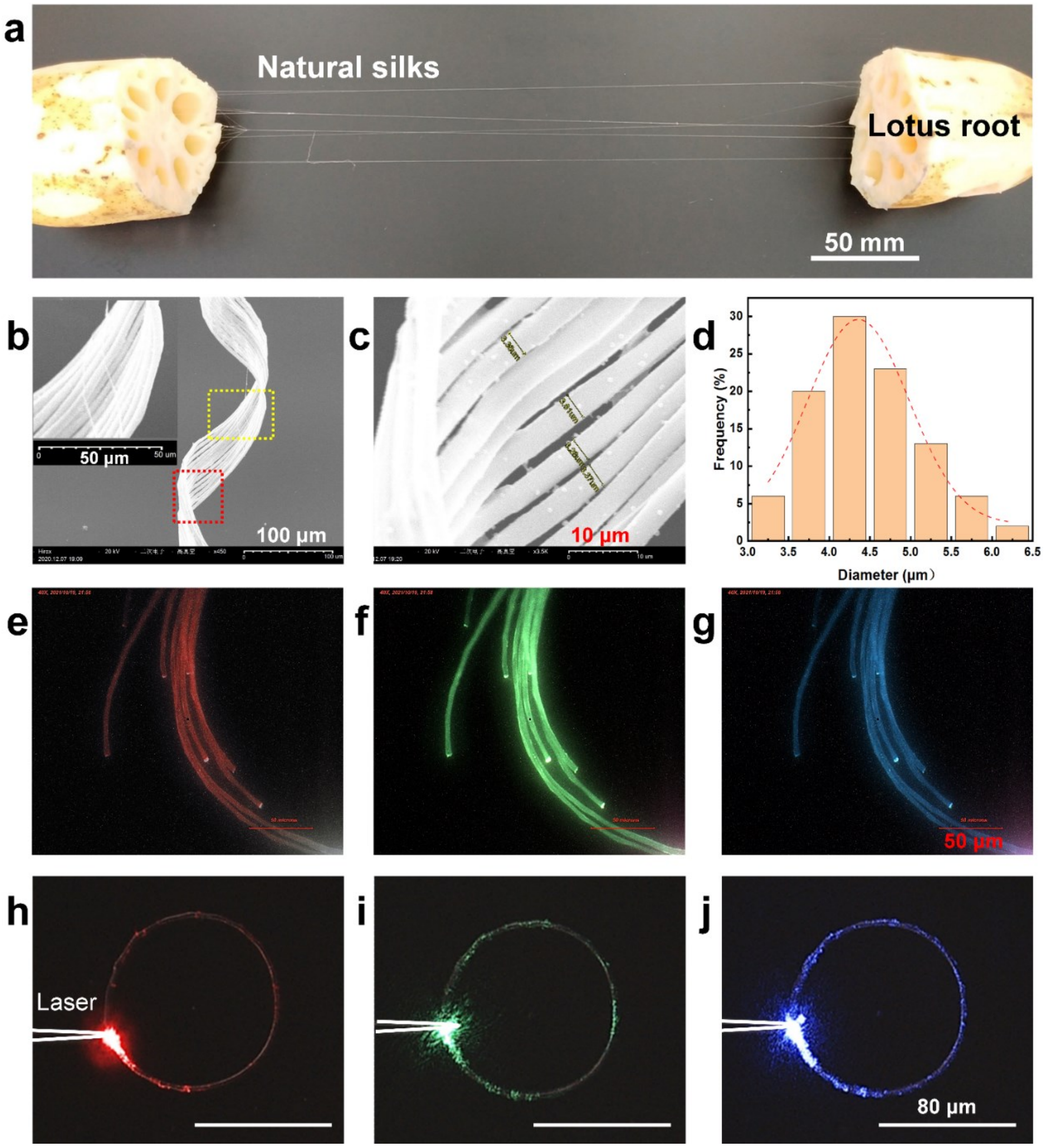

**Figure 1. Microstructure of single-filament microfibers.** (a) Photograph of natural silks directly extracted by hands from lotus root. (b) SEM micrographs of micro-silks from lotus root. Inset: magnified view of the yellow dashed region. (c) Magnified SEM micrograph of micro-silks highlighted by the red dashed line in (b). (d) Diameter distribution of the filaments. Average ~ 4.25 µm, standard deviation ~ 0.04 µm. (e-g) Fluorescence micrographs of microfibers, acquired through various long-pass filters to attenuate the excitation light at 532 (e), 440 (f), and 365 nm (g), respectively. Scale bars in (e-g): 50 µm. (h-j) Optical micrographs of a single-filament microfiber with evanescently-coupled light from a tapered fiber (left side of the images). The coupled light has wavelengths of 655 (h), 532 (i) and 440 (j) nm, respectively. The micrographs are acquired through short-pass filters. Scale bars in (h-j): 80 µm.

The broadband emission of a single filament is peaked at a wavelength of about 453 nm, and it is stable along the microfiber length (Figure 2a). The average fluorescence lifetime is about 1.85 ns (Figure 2b), which also does not vary significantly along the microfiber length. Instead, a red-shift by 20-30 nm is found in the fluorescence spectrum by changing the excitation wavelength from 458 nm to 514 nm (Figure S5). To investigate the origin of the observed emission, the composition of the single-filament microfibers is investigated by both energy dispersive X-ray spectroscopy (EDS) and micro-Raman spectroscopy (Figures 2d, S6 and S7). Bands characteristic of cellulose and lignin are identified in the micro-Raman spectra such as, for instance, the band at 890-913 $cm^{-1}$ attributed to β glycoside bonds, HCC and HCO bending and C-O-C symmetric stretching of cellulose [30, 31, 32]. The bands in the interval 1000-1200 $cm^{-1}$ are assigned to HCC, HCO and HOC bending of cellulose [30-33], whereas the band at about 1331 $cm^{-1}$ is attributed to lignin [34]. The bands around 1590–1600 $cm^{-1}$ and 1650–1660 $cm^{-1}$ are assigned to the ring and C=C stretching of lignin [32, 35], the bands at about 2890–2910 $cm^{-1}$ to CH and $CH_2$ stretching of cellulose [30, 33], and the characteristic peaks at 3370–3390 $cm^{-1}$ to the stretching of OH– groups of cellulose I [30, 32]. The composition of the microfibers is quite homogeneous through their length (Figure S7). Fluorescence has been reported for both cellulose and lignin in the visible spectral range, as a broad emission band peaked in the interval 400-500 nm depending on the optical excitation conditions, the source of the raw material and the preparation method [36-38], and with lifetimes of 0.2-5 ns [39]. Moreover, a red-shift of the emission upon reducing the excitation energy has also been observed in lignin samples [40-42]. Overall, these results suggest a high versatility of the fluorescent, Lotus-derived micro-fibers, being easily tailorable by proper optical excitation and potentially controllable at high modulation rates (up to GHz).

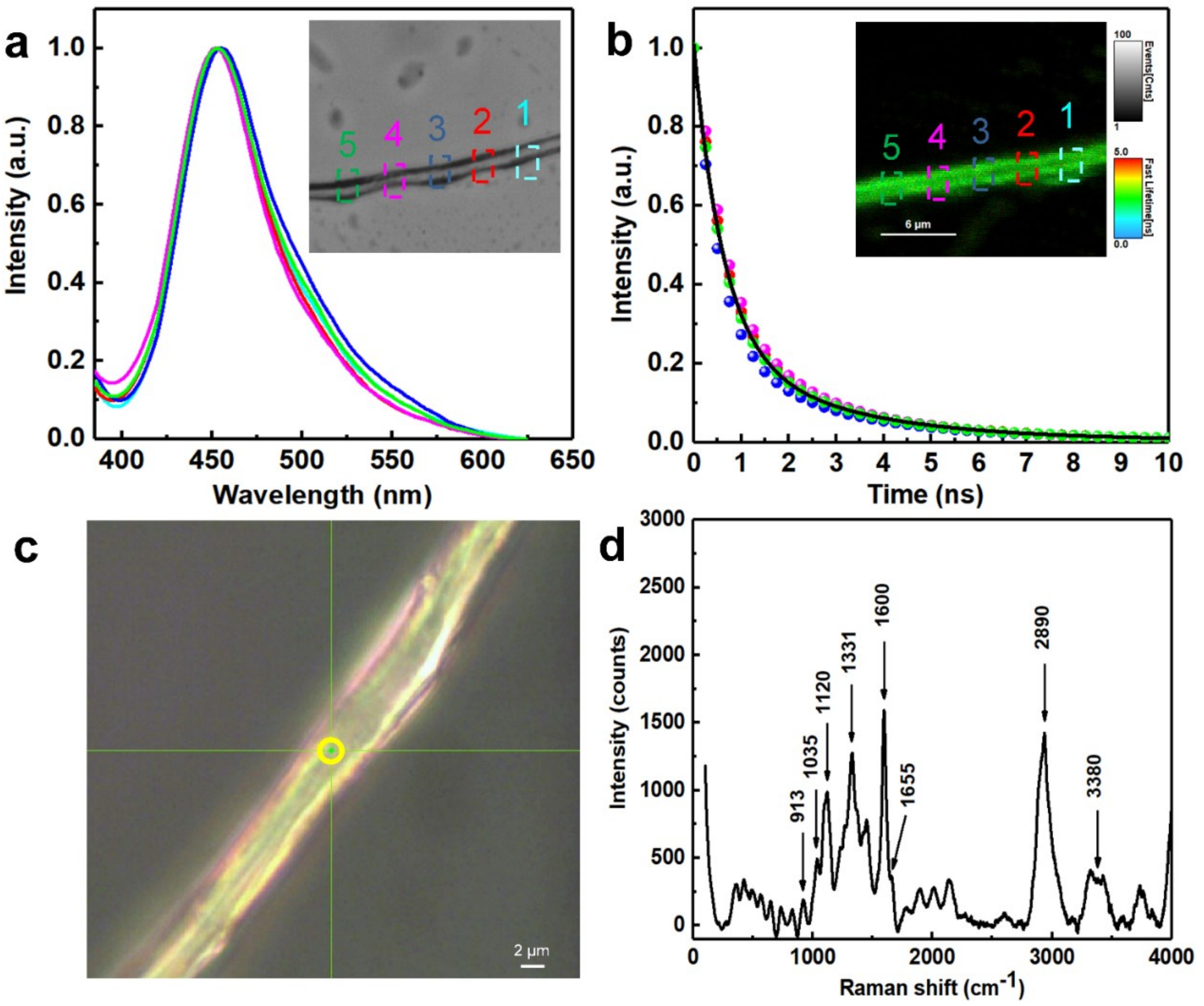

**Figure 2. Fluorescence and Raman spectra of single-filament microfiber.** (a) Confocal fluorescence spectra collected from five areas along the length of a microfiber (dashed boxes in the micrograph shown in the inset). (b) Time-resolved fluorescence curves collected in various areas of a microfiber (dashed boxes in the micrograph shown in the inset). The continuous line is a fit to the data by an exponential function. The microfiber is excited by 405 nm ps-pulsed light. (c) Optical micrographs of a single-filament microfiber used for micro-Raman measurements. (d) Micro-Raman spectrum of a single-filament microfiber.

The optical properties of lotus fibers make them suitable as both passive and active waveguides. These types of waveguides mainly differ in the sources of the light coupled in them [43-46]: while in active waveguides the guided light is generated through the photo-excited fluorescence of the material that constitutes the waveguide, in passive waveguides light must be coupled into the waveguide from an external light source. Figure 3a-c are micrographs of a single-filament microfiber in which lasers of 655, 532 and 440 nm, respectively, are launched in a tapered optical fiber and evanescently-coupled into the lotus-derived passive waveguide. Another tapered optical fiber is evanescently coupled with the microfiber end tip, for accurate detection of the output optical power. Figure 3d-f clearly show that laser light well-couples into the single-filament microfiber, and is then guided along the filament length. The surface scattering of light transported along the microfiber body is unnoticeable, and much weaker than that at the input and output regions where bright spots highlight effective waveguiding [47]. The propagation loss coefficient, $\alpha$, of the single-filament microfiber with 655, 532, and 440 nm lasers is calculated as equation (1):

$$\alpha = \frac{-10\log\frac{P_{out}}{P_{in}}}{L} \tag{1}$$

where $P_{out}$ and $P_{in}$ are the optical power at the output of the microfiber and at the input tapered fiber, respectively, and $L$ is the propagation distance. Propagation loss are higher for smaller wavelengths, being of 0.160, 0.144, and 0.138 dB/μm for blue, green, and red light, respectively (Figure 3g). These results are in agreement with numerical simulations (see Figure S8).

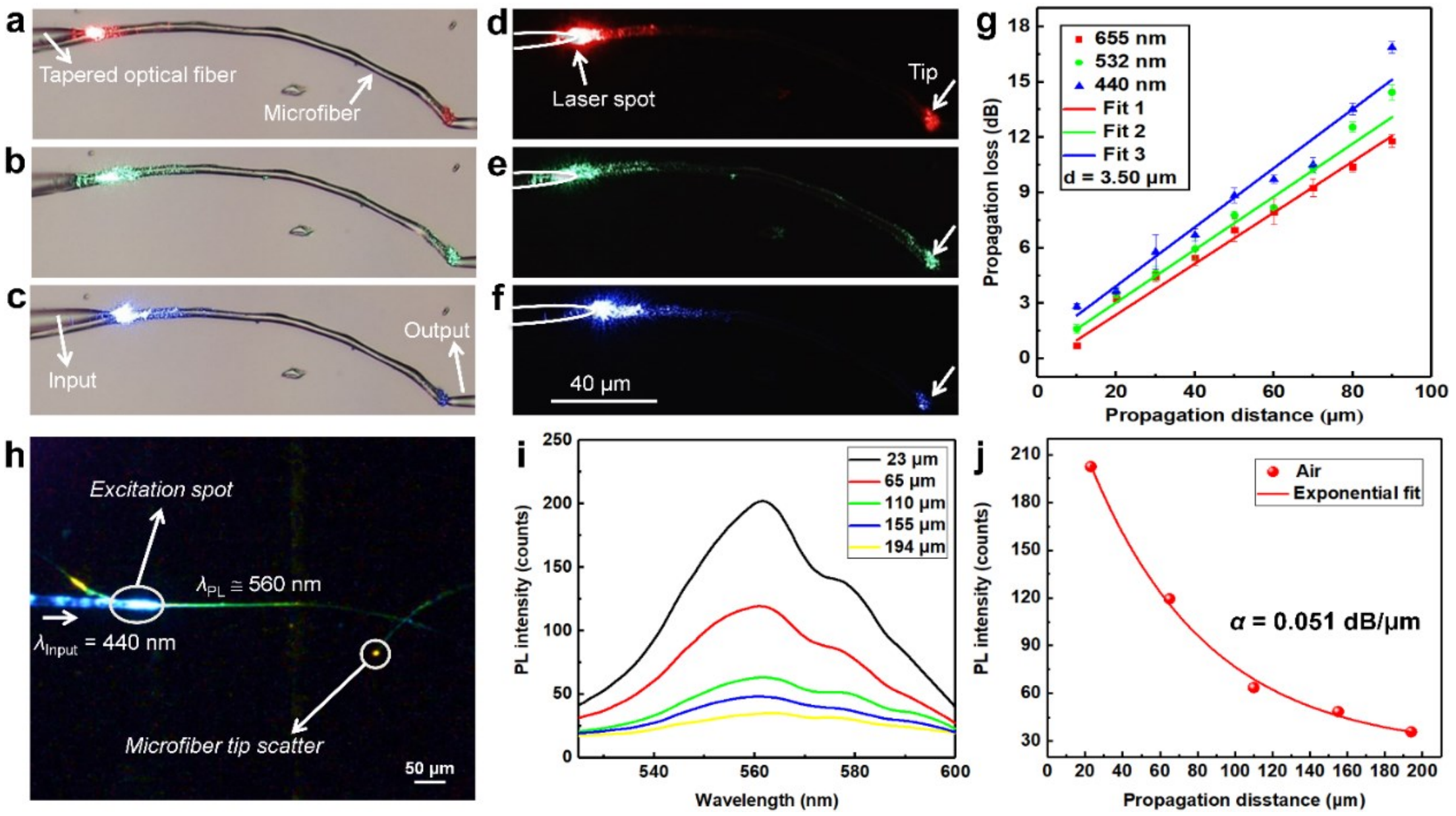

**Figure 3. Passive and active waveguiding in single-filament microfiber.** (a-c) Bright-field optical micrographs of a single-filament microfiber (3.5 μm diameter) coupled to 655, 532 and 440 nm lasers, at a power of 50 nW. (d-f) Corresponding dark-field optical micrographs. Bright spots can be seen at the two terminations of the single-filament microfiber. Scale bar in (f) is applicable to (a-e). (g) Propagation loss *vs*. propagation distance in the passive waveguide. (h) Dark-field optical micrograph of a single-filament microfiber acting as active waveguide. Color filters are not used. The 440 nm excitation laser leads to 560 nm fluorescence which is then waveguided along the microfiber. Another microfiber in crossed geometry is lighted up, resulting in bright tip scattering. (i) Intensity of the fluorescence signal at different propagation distances. (j) Fluorescence intensity at 560 nm *vs*. propagation distance. The continuous line is a fit to the data by an exponential function.

The single-filament microfibers might also operate as active waveguides, transporting their own fluorescence signal (Figure 3h). By coupling 440 nm light into one end of a single-filament microfiber, one can observe the fluorescence at different propagation distances, as shown in Figure 3i. From the analysis of the fluorescence intensity at different propagation distances, we obtain that the propagation loss coefficient of the active waveguide is 0.051 dB/μm (Figure 3j). In addition, Figure 3h show that the single-filament microfiber can also light up other branched microfibers in crossed geometry.

Propagation loss of these waveguides, particularly in passive operation, are relatively large compared to other materials [48-51], however appealing and effective sensing schemes can be envisaged by the natural, biocompatible and highly flexible filaments. When light propagates in the single-filament microfiber, part of it extends in the surrounding environment through evanescent fields. The single-filament microfiber is therefore sensitive to various classes of changes in the surrounding environment [52]. Figure 4a show a schematics of pH sensing based on the active waveguides. To this aim, a microchannel is constructed through two cover slips on a silica substrate. The single-filament microfiber is suspended on the micro-channel by a conical fiber, and both ends are fixed with glue bonding. Figure 4b show the SEM micrograph of the 4 μm diameter microfiber used for sensing. Figure 4c, d show the 300-μm-long microfiber placed on the 200-μm-width micro-channel without (c) and with (d) 440 nm laser excitation, respectively. Both ends of the micro-channel are opened to facilitate the flow of buffer solutions with different pH. After the microfiber was completely immersed in a buffer solution for 5 min, the 440 nm laser is coupled into the left end of the microfiber (spot A in Figure 4a). The 440 nm laser propagates in the microfiber and excites fluorescence. The fluorescence spectrum is measured at spot B shown in Figure 4a. Figure 4e show the optical micrographs of the single-filament microfiber for different pH of liquid microenvironments. We can measure the photoluminescence (PL) intensity of the microfiber in different pH of liquid microenvironments and at different sensing lengths (Figure 4f). The optimal sensing length of the pH sensing is found to be at 65 μm (see Figure S9). Upon analyzing the photoluminescence (PL) intensity of the microfiber at different propagation distances and for buffer solutions with different pH (see Figure S10), we summarize the propagation loss coefficient values in Figure 4g. The observed trend could be in part related to the hydrolysis of single-filament microfiber in an acidic environment. The main component of the single-filament microfiber is cellulose (Figure 2d and S7) [32]. Cellulose side bonds undergo catalyzed hydrolysis in an acidic environment to generate a negatively charged aldehyde group, thus forming a shell on the surface of the microfiber and possibly leading to larger quantum yields and fluorescence intensity [53]. Figure 4h show PL spectrum of the microfiber in buffer solutions with different pH at the optimal sensing length. The PL intensity decreases with increasing solution pH, that is, the fluorescence intensity of the microfiber is larger in acidic environment and smaller in alkaline environment. In acid environments, the increase of cations can lead to the charge-charge repulsion, structural stiffening and consequently higher emission. Such fluorescence response is also consistent with intrinsic clusterization-triggered emission [44, 54, 55]. The variation of the fluorescence intensity with pH is reversible, as found by measuring the PL intensity upon increasing the pH from 3 to 9 and vice versa in liquid microenvironment, for ten consecutive cycles (Figure S11).

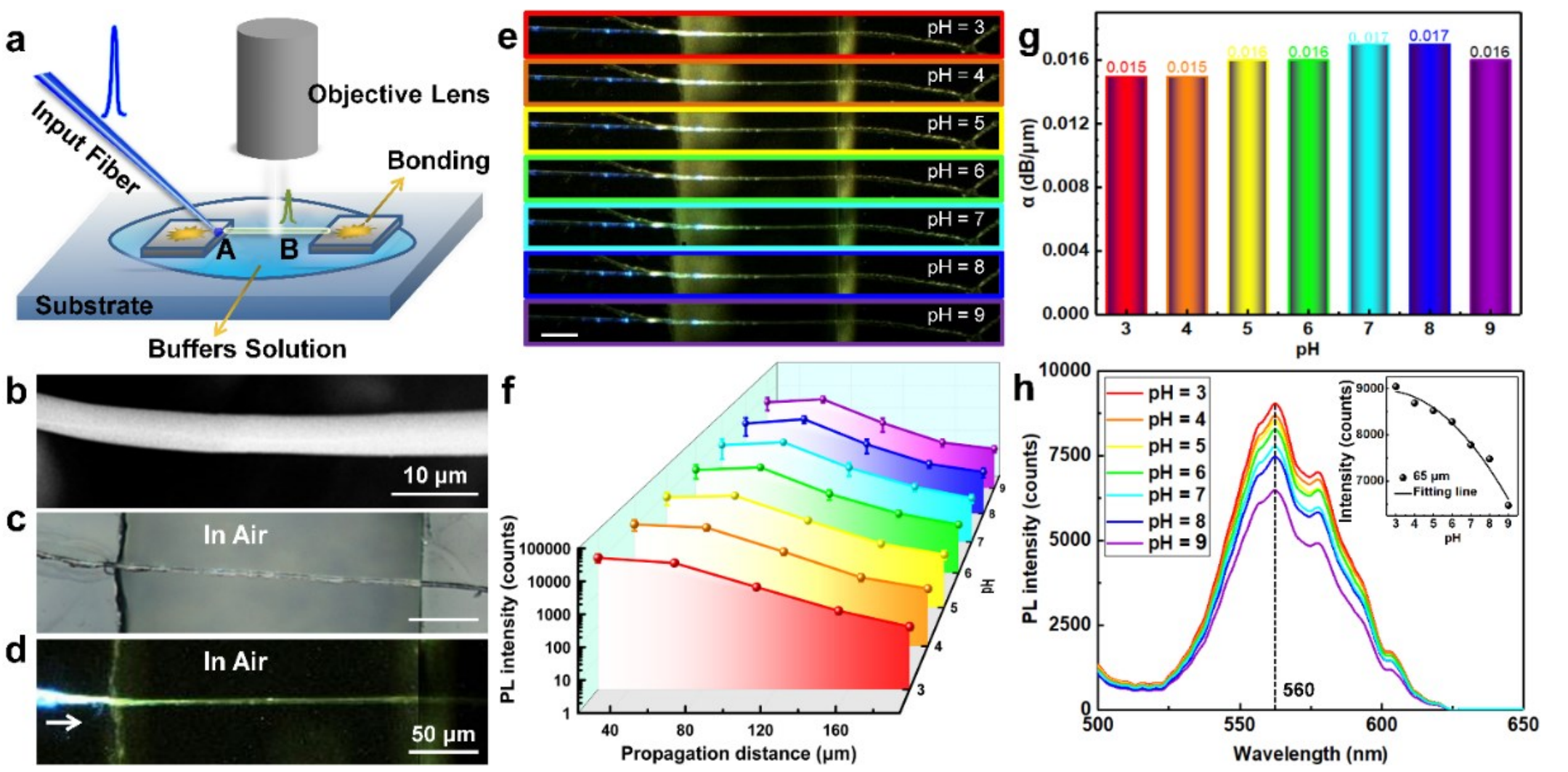

**Figure 4. Optical sensing of pH with single-filament microfiber.** (a) Scheme of the system developed for pH sensing. The microfiber is immersed in buffer solutions with different pH. The sensing length is the distance between the excitation spot, A, and the area of collection for the fluorescence, B. (b) SEM of the single-filament microfiber used in the experiment. (c, d) Bright-field (c) and dark-field (d) optical micrographs of a single-filament microfiber suspended through the microchannel. Power of the 440 nm laser excitation: 100 nW. (e) Dark-field optical micrograph of the microfiber device immersed in liquid microenvironments with different pH. Scale bar: 50 μm. (f) PL peak intensity as a function of propagation distance (25, 65, 105, 145, and 185 μm) and pH. (g) Active waveguide loss coefficient as a function of pH. (h) PL spectra of the microfiber monitor (65 μm sensing length) with different surrounding pH. Inset: PL intensity at 560 nm, *vs*. pH.

Furthermore, the single-filament microfibers can be used as passive waveguides to monitor bacteria cell apoptosis (Figure 5a). The bacteria used in this experiment is *Helicobacter Pylori*, one of the causes of human stomach diseases [56]. Light (440 nm wavelength and input power of 10 nW) is coupled to the left end of the microfiber schematized in Figure 5a by a tapered fiber (fiber 1 in Figure 5a), and the output power is collected with a second tapered fiber (fiber 2 in Figure 5a) at the right end of the microfiber. $H_2O_2$ is used to induce bacterial apoptosis. Taking the 3.5 μm-diameter single-filament microfiber with a sensing length of 65 μm as an example, Figure 5b shows the optical micrograph of the single-filament monitor in buffer solution at initial time (*t* = 0), that is the time at which $H_2O_2$ is added to the bacteria suspension. After 25 min (Figure 5c), the activity of *Helicobacter Pylori* gradually decreases. After 50 min (Figure 5d), the bacterial cells are reduced to a dispersed particulate attaching to the surface of the single-filament microfiber. As a consequence, an increase of the light scattered through the fiber body occurs, resulting in a decrease of the intensity of light transmitted through the micro-fiber. Therefore, the either healthy or apoptotic (Figure 5e) status of *Helicobacter Pylori* (see Figure S12) can be monitored by detecting the output power of the single-filament microfiber end (Figure 5f, red data). Upon apoptotic events, the output light intensity decreases with the increasing time, representing the increase of the surface scattering caused by apoptotic *Helicobacter Pylori* attached to the microfiber surface. After 45 min, the light intensity remains stable, which means that almost all *Helicobacter Pylori* in the solution have reached an apoptotic state. For comparison, a healthy *Helicobacter Pylori* solution (no $H_2O_2$ added) with the same *Helicobacter Pylori* concentration is prepared and monitored in the same way. The time-dependent output light intensity for the healthy *Helicobacter Pylori* is shown in black data of Figure 5f. With the increase in time, the light intensity fluctuates slightly, meaning that the *Helicobacter Pylori* without $H_2O_2$ can maintain good activity within 50 min. No significant variation of the intensity of the light transmitted by the microfiber is observed after addition of $H_2O_2$ in a solution without bacteria (Figure 5f). To corroborate the experimental findings, Raman spectroscopy of the *Helicobacter Pylori* without/with $H_2O_2$ at different times is carried out as parallel test of undergoing apoptosis processes of the *Helicobacter Pylori*, as shown in Figure 5g, h. For the *Helicobacter Pylori* without $H_2O_2$, a stable Raman peak is measured at 1378 $cm^{-1}$ (Figure 5g). In Figure 5h, there are instead two Raman peaks at 876 and at 1378 $cm^{-1}$, with the 876 $cm^{-1}$ peak resulting from the presence of $H_2O_2$ [57, 58]. Here the intensity of the Raman band at 1378 $cm^{-1}$ decreases upon increasing time, indicating the increase of the apoptosis degree (Figure 5h). These results can be explained by considering that in a typical process of bacterial cell apoptosis, changes in the permeability of the bacterial cell membrane lead to the gradual release of enzymatic substances from the bacterial cell interior into the surrounding liquid microenvironment [15, 59]. Such compounds may adhere to the single-filament microfiber surface, increasing surface roughness or local thickness. This leads to an increase of the scattering of light propagating through the microfiber and a decrease of the output light intensity, as here observed (Figure 5f) and reported in similar experiments performed with polymer nanofibers [15].

Overall, data of Figure 5 show that the apoptosis of *Helicobacter Pylori* can be monitored in real time by the single-filament natural and eco-friendly microfiber. In perspective, combining the passive and active waveguide properties, one could simultaneously detect the pH change and monitor bacterial activity by using the same microfiber.

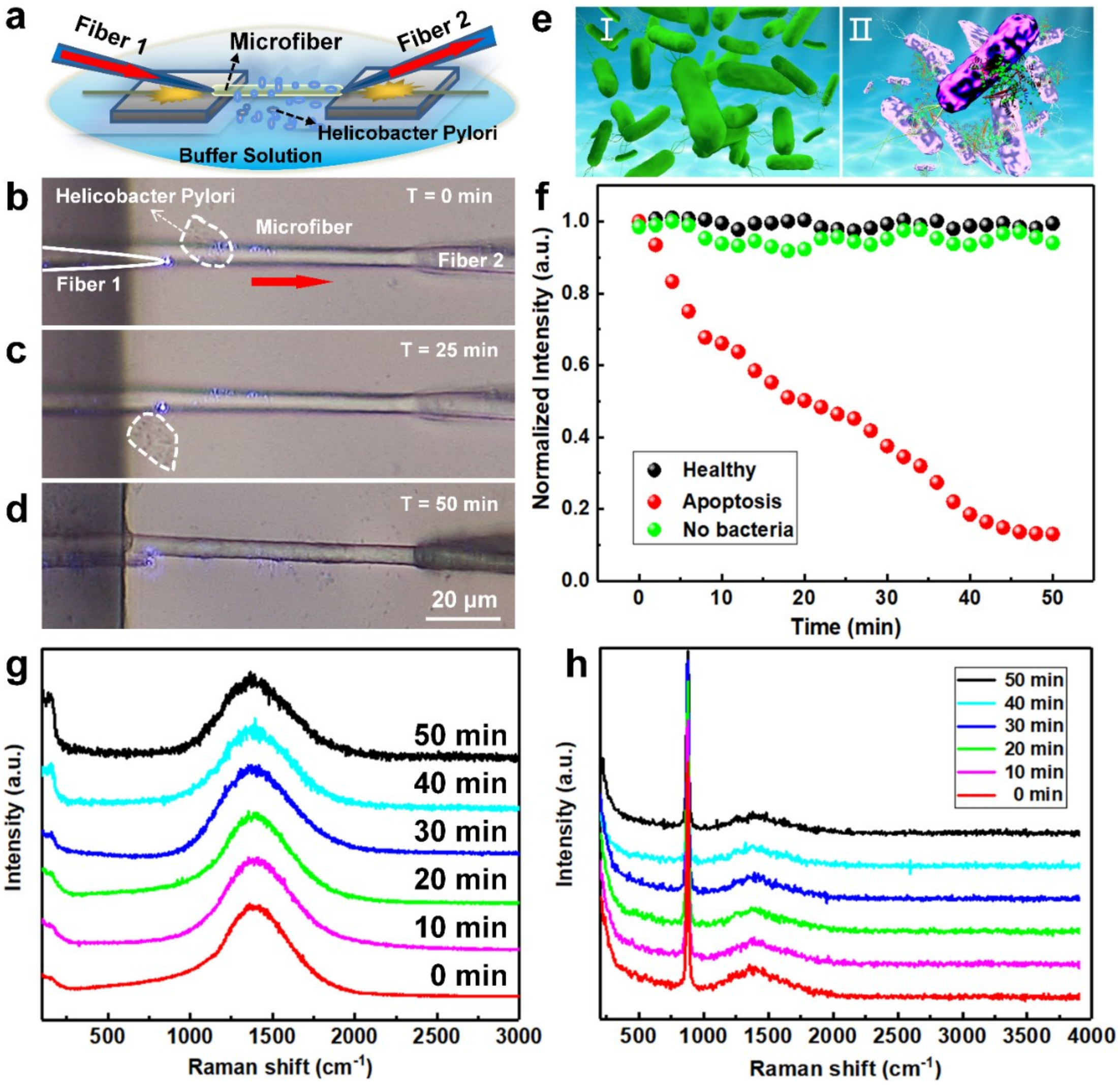

**Figure 5. Monitoring the activity of *Helicobacter Pylori* with single-filament microfiber.** (a) Schematic diagram of the system developed for the real-time monitoring of the activity of *Helicobacter Pylori*. (b-d) Optical micrograph of the microfiber after 0, 25 and 50 min of operation. 440 nm laser light (10 nW) is coupled into the microfiber from the input fiber 1, while 30% $H_2O_2$ is added at $t$ = 0 min. The red arrow represents the direction of light propagation. (e) Schematic illustration of the *Helicobacter Pylori* in healthy conditions (I) and apoptosis (II). (f) Time-dependent output light intensity for apoptotic (red data, with $H_2O_2$) and healthy (black data, without $H_2O_2$) *Helicobacter Pylori*. Data collected by adding $H_2O_2$ in a buffer solution without bacteria are shown as green circles. (g, h) Raman spectra from $t$ = 0 to 50 min for the healthy (g) and apoptotic (h) *Helicobacter Pylori*.

In summary, a novel biosensor scheme based on a single-filament microfiber with waveguiding and light-emitting properties is successfully prepared through physical separation of natural lotus silks. The single-filament microfibers have excellent full-color fluorescence properties, exploitable for pH monitoring, and can be used as biosensor to enable real-time monitoring of bacterial activity. Compared to polymer fibers such as polylactic acid [15, 60], the single-filament microfibers directly extracted from natural lotus root has the advantages of simple production, low cost and low environmental impact. Moreover, compared with other metal or semiconductor microfibers [61, 62], single-filament microfibers might have better biocompatibility, biodegradability and mechanical flexibility. They can be an appealing platform for detecting the properties of liquid environment and the activity state of living cells, and for the development of eco-friendly photonics for biophysics and biomedicine.

## Experimental

**Fabrication of the microfiber.** The commercial mature rhizomes of lotus (*Nelumbo nucifera Gaertn*) were purchased from a market near Jinan University and used as starting materials. Using manual extraction methods to obtain primary micro-silks from lotus root. To separate the single-filament microfibers from multi-filaments micro-silks, a chemically-assisted physical separation method was used. First, we soaked the multi-filaments in alkali solution, the single-filament gradually disperse and gaps appear between filaments. This is due to pectin and other components between filaments being removed by the alkali solution [63]. Next, the as-treated multi-filaments were transferred to clean glass slides, washed repeatedly with deionized water to remove the residual alkali solution, and then fixed on one end with glue. Finally, the separation of single-filament microfiber was achieved by micromanipulation with two conical fibers under optical microscopy.

**Preparation of the buffer solutions.** 0.9 mL of hydrochloric acid were slowly poured into 1000 mL of water to obtain 0.01 mol/L of hydrochloric acid. Afterwards, it was diluted 10 times sequentially to obtain a buffer with pH = 3-6. Next, 0.4 mg of NaOH particles were dissolved in 1000 mL of water to obtain a buffer with pH = 9, and then diluted 10 times to obtain a buffer with pH = 8. Finally, the dilute NaOH solution was titrated with dilute hydrochloric acid to obtain solutions with pH = 7. The pH was measured with a pH meter.

**Characterization.** The morphology of the microfibers was characterized with a Hirox-SH-5000M SEM operating at an accelerating voltage of 20 kV. Chemical elements of the microfibers were examined by the EDS using a Bruker AXS Quantax system working at 15 kV. Fluorescence spectra were measured with a Zeiss CRAIC 20/30 PV$^{TM}$ microspectrophotometer through a 40× objective with a sampling spot of 1.0-5.0 μm in diameter. The microfibers were micro-manipulated with a 3D video microscopy through a 50× objective.

**Raman and fluorescence lifetime characterization.** Raman characterization was performed under backscattering geometry with 532 nm laser excitation. Raman characterizations of *Helicobacter Pylori* without/with $H_2O_2$ at different times were performed with a laser confocal microscope (Jobin Yvon Horiba, Xplora) through a 50× objective (NA = 0.50). The spectra were recorded from 100 to 4000 cm$^{-1}$ spectral range using a low frequency filter, 1200 grooves per mm grating, and Peltier cooled charge coupled device. Lifetime mapping were recorded by confocal microscope system (PicoQuant, MicroTime 200). Excitation power of 405 nm pulsed laser was controlled by a PDL-800B driver (PicoQuant). The lifetime spectra were integrated for 30 s. PL analysis of various microfibers was performed by confocal fluorescence microscopy by a Leica TCS SP2 laser scanning head coupled to an inverted microscope (Leica DMIRE2). Three different excitation wavelengths were used for optical excitation: 458 nm, 488 nm and 514 nm. The excitation lasers were focused onto the sample through a 20× objective (Leica, NA = 0.5). The sample emission was collected through the same objective used for the excitation and the fluorescence signal was detected by a photomultiplier tube.

**Numerical simulations.** The electromagnetic energy density was obtained by using a collection of online resources (www.computational-photonics.eu). Numerical simulations of microcavity optical modes were performed to study whispering gallery modes of circular 2-D dielectric optical cavities in microfiber section. FiMS modes of circular multi-step index optical fibers and 1-D mode solver for waveguide bends were also studied [64]. Among them, the cross-section field distribution mainly selects the TE modes [65]. The effective real part of the complex refractive index of the single-filament microfiber was measured by the Abbe refractometer. The average results of 1.495 are close to the cellulose nanocrystals at the corresponding wavelengths [66].

## Acknowledgements

This work was supported by the National Natural Science Foundation of China (Nos. 11804120 and 61827822) and the Guangdong Basic and Applied Basic Research

Foundation (2023A1515030209), the Research Projects from Guangzhou (2023A03J0018), the Fundamental Research Funds for the Central Universities (21623412). D.P. acknowledges the funding from the Italian Minister of University and Research PRIN 2017PHRM8X project ("3D-Phys").

## Supporting Information

Surface roughness, flexibility, simulated energy density, SEM, EDS, Micro-PL and micro-Raman characterizations.

# Supporting Information

# Light-Emitting Microfibers from Lotus Root for Eco-friendly Optical Waveguides and Biosensing

Xianguang Yang[1]*, Liping Xu[1], Shijie Xiong[1], Hao Rao[1], Fangchang Tan[1], Jiahao Yan[1], Yanjun Bao[1], Annachiara Albanese[2], Andrea Camposeo[3], Dario Pisignano[2,3], and Baojun Li[1]

[1]Guangdong Provincial Key Laboratory of Nanophotonic Manipulation, Institute of Nanophotonics, Jinan University, Guangzhou 511443, China

[2]Dipartimento di Fisica, Università di Pisa, Largo B. Pontecorvo 3, I-56127 Pisa, Italy

[3]NEST, Istituto Nanoscienze-CNR and Scuola Normale Superiore, Piazza S. Silvestro 12, I-56127 Pisa, Italy

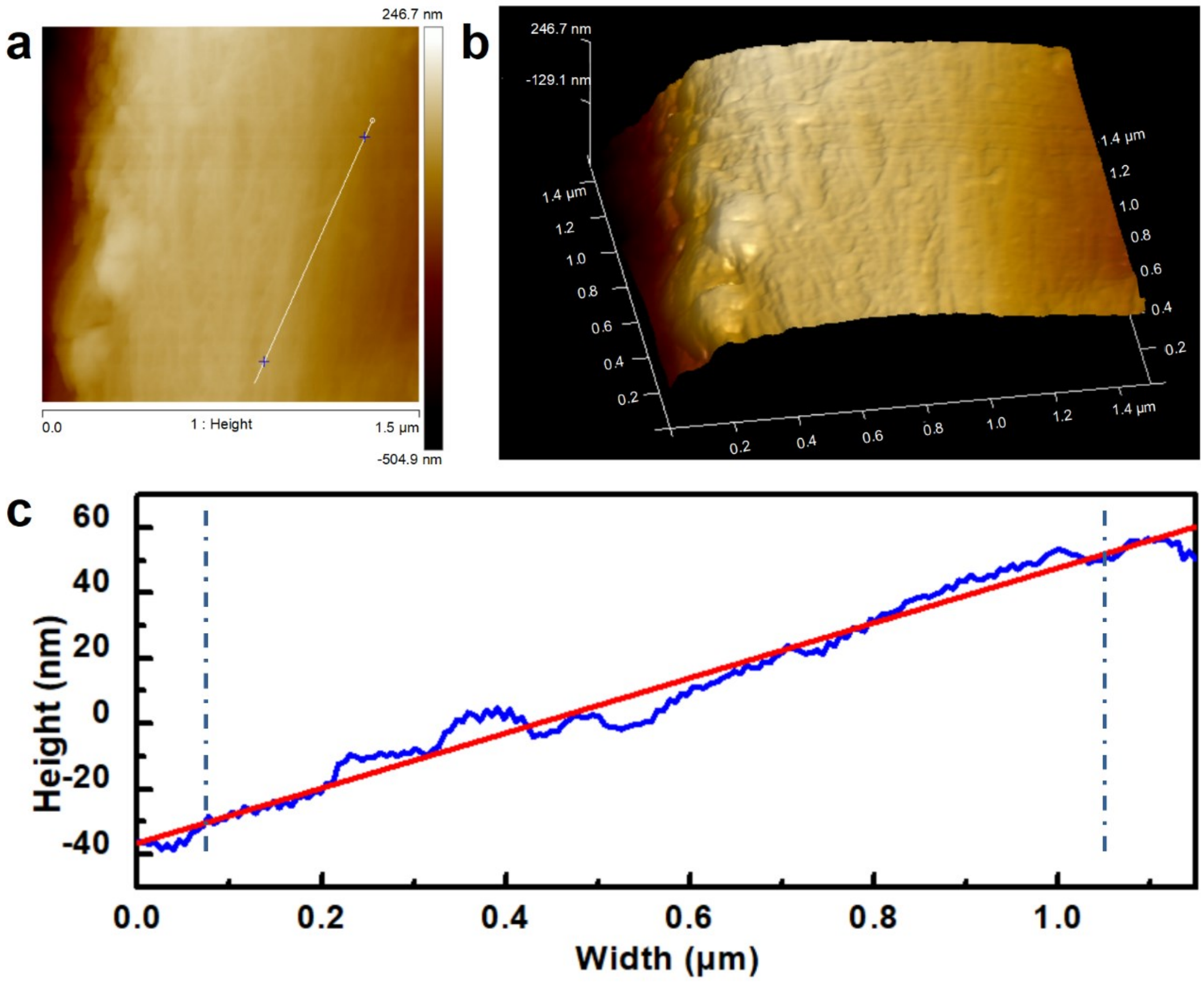

**Figure S1**. **Characterization of the surface roughness of single-filament microfiber**. (a) Atomic force microscopy image of a single microfiber surface. Size of the imaged area: 1.5×1.5 μm$^2$. (b) Three-dimensional height topography of a single microfiber surface. (c) The blue curve is the height profile along the white line in (a). The red line is the fitted reference line of the surface roughness, from which a root-mean-square surface roughness of about ~4 nm is estimated.

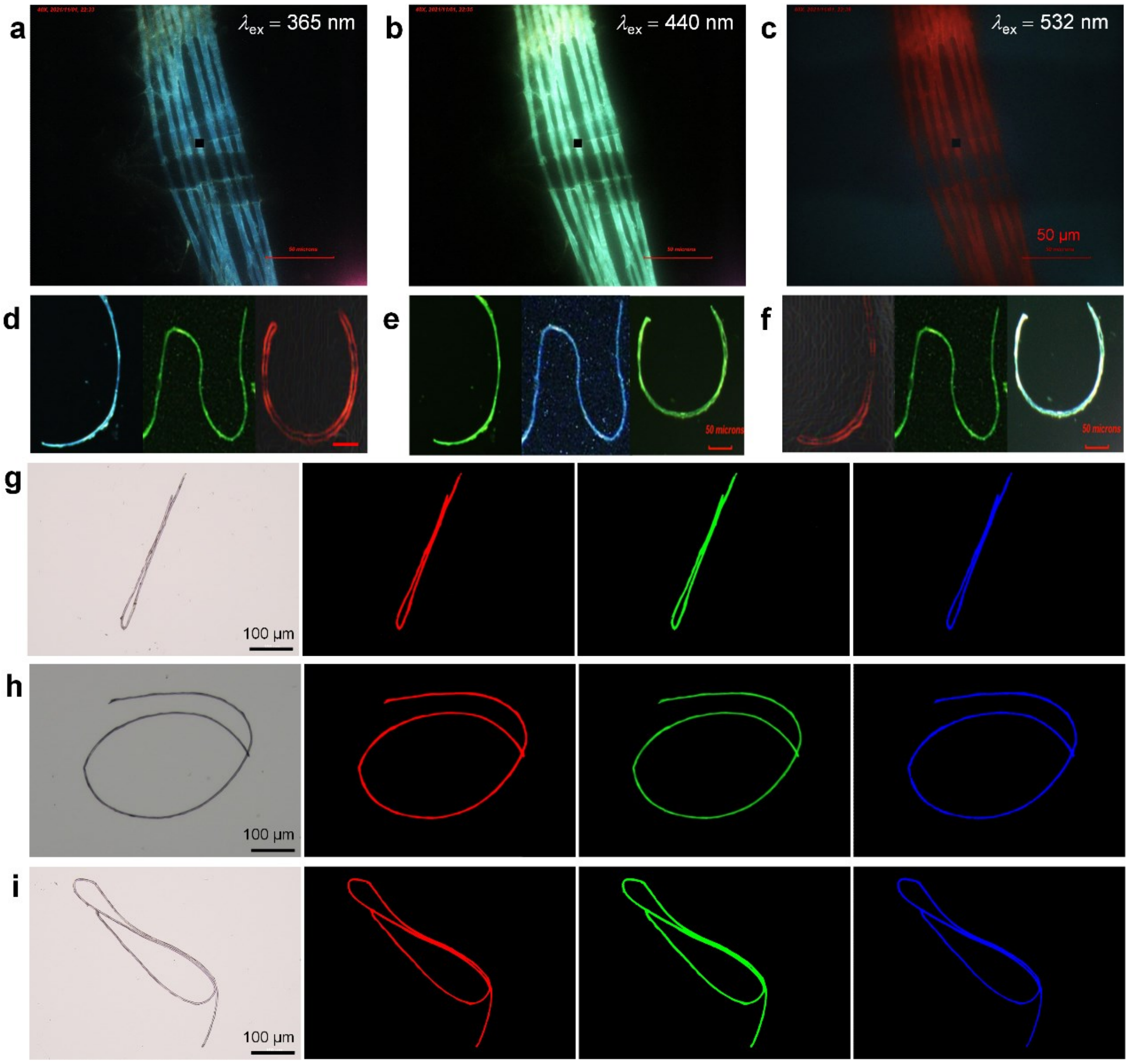

**Figure S2. Fluorescence characterization of robust single-filament microfibers.** (a-c) Fluorescence micrographs of a beam of micro-silks under the excitation of 365, 440, and 532 nm lasers, respectively. (d-f) Single-filament microfibers can be manipulated to make different letters of "JNU". Prototype demonstration for potential applications in full-color micro-display under the excitation of different lasers. (g-i) Bright field (left image) and fluorescence micrographs (middle and right images) of single-filament microfibers, which were shaped into different complex geometries.

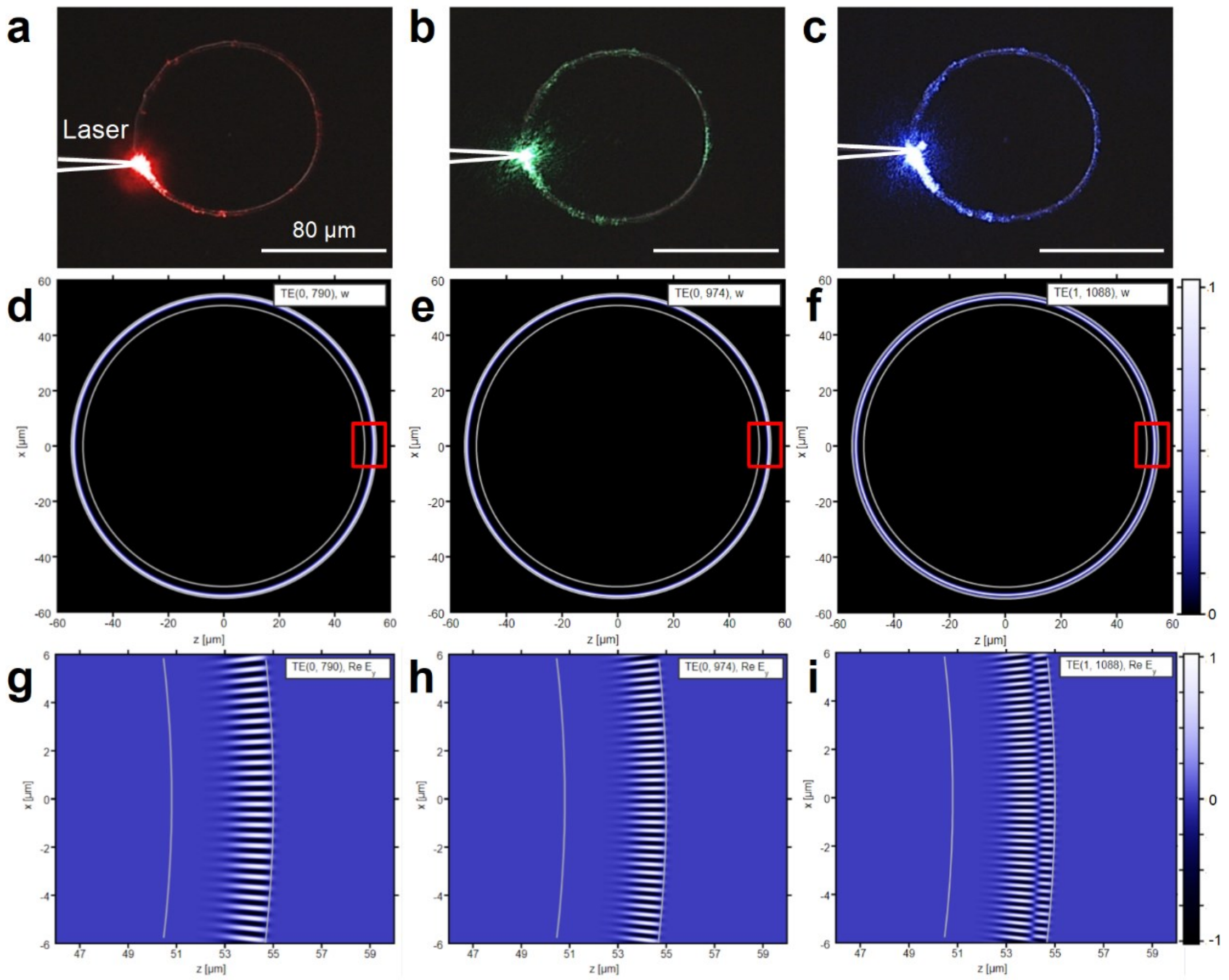

**Figure S3.** Simulated electromagnetic energy density (d-f) and electric field (g-i) distribution of the whispering gallery mode with $TE_{(x,\ y)}$ polarization in the microfiber loop cavity (experimental micrographs in a-c, respectively). Red line in (d-f) indicates the area shown in (g-i).

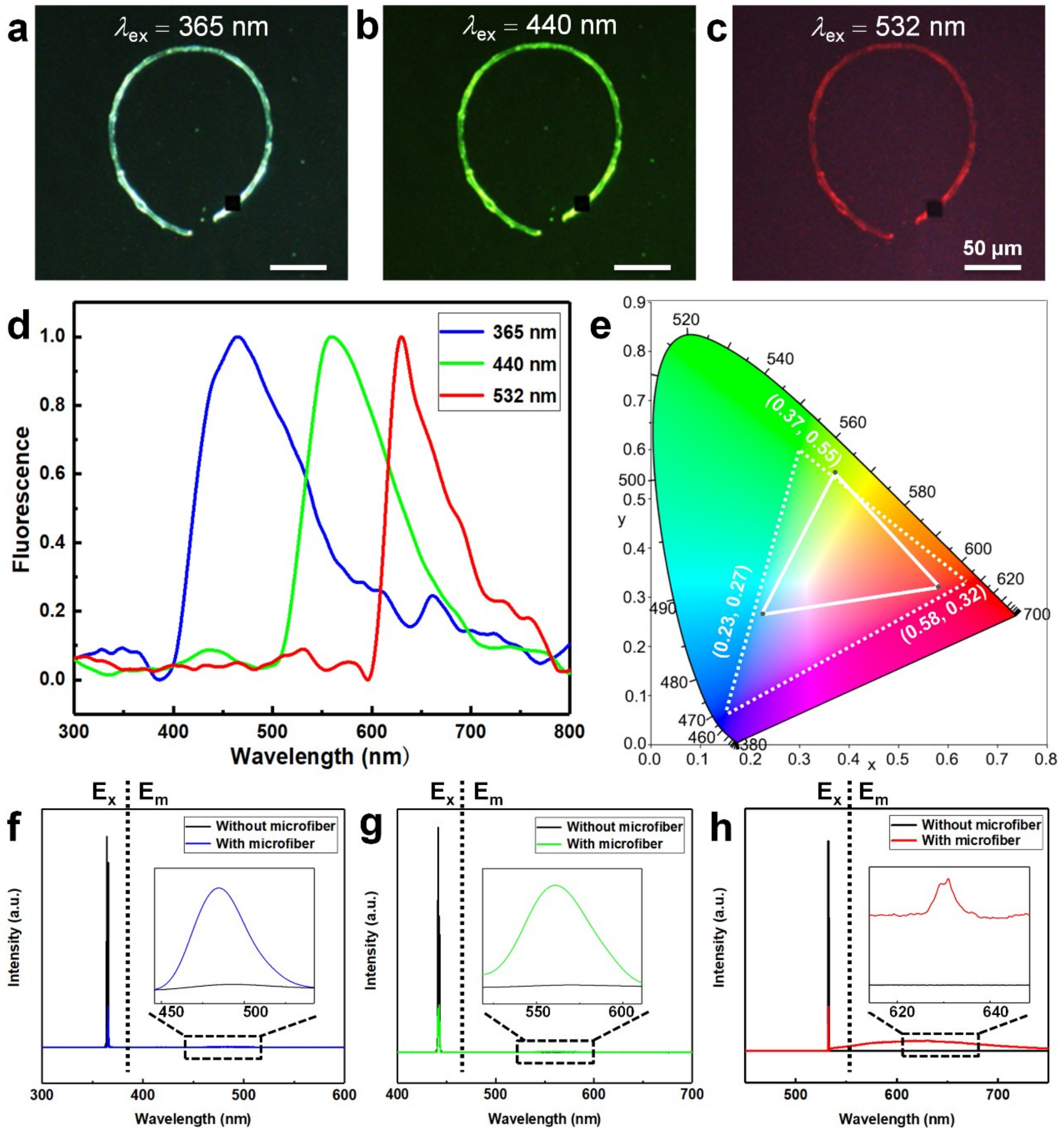

**Figure S4. Broadband fluorescence characterization of robust single-filament microfiber.** (a-c) Fluorescence micrograph of the same single-filament microfiber excited with 365 (a), 440 (b), and 532 (c) nm lasers in an open-ring shape. Scale bar in (c) is applicable to (a, b). (d) Normalized fluorescence spectra of the same microfiber excited by 365, 440 and 532 nm laser illumination. (e) Color coordinates of red, green and blue emissions from lotus microfiber. Dashed line shows the standard RGB. (f-h) The fiber emission efficiency of the red (c, h), green (b, g), and blue-emissive (a, f) single-filament microfiber.

Figure S4e show the obtained color coordinates of red fluorescence (0.58, 0.32), green fluorescence (0.37, 0.55), and blue fluorescence (0.23, 0.27), compared to the three primary color coordinates of R (0.64, 0.33), G (0.30, 0.60) and B (0.15, 0.06). The broadband fluorescence area rate of the single-filament microfiber was calculated to reach at 42.08% of the triangle area from industrial RGB, indicating high potential capability in full-color displays with wavelength-converted luminescence [1]. Figure S4f-h show the fiber emission efficiency of the red-, green-, and blue-emissive single-filament microfiber. The efficiency can be calculated according to the equation (1):

$$\eta = \frac{I_{em}}{I_{ex,no\ microfiber} - I_{ex,with\ microfiber}} \tag{1}$$

where $I_{em}$ is the integrated area of the emission spectra of the microfiber, $I_{ex,no\ microfiber}$ and $I_{ex,with\ microfiber}$ are the integrated area of the excitation lasers at 365, 440 and 532 nm without and with microfiber, respectively.

Thus, the $\eta$ was calculated to be 18%, 20%, and 14% for red-, green-, and blue-emissive single-filament microfiber, respectively.

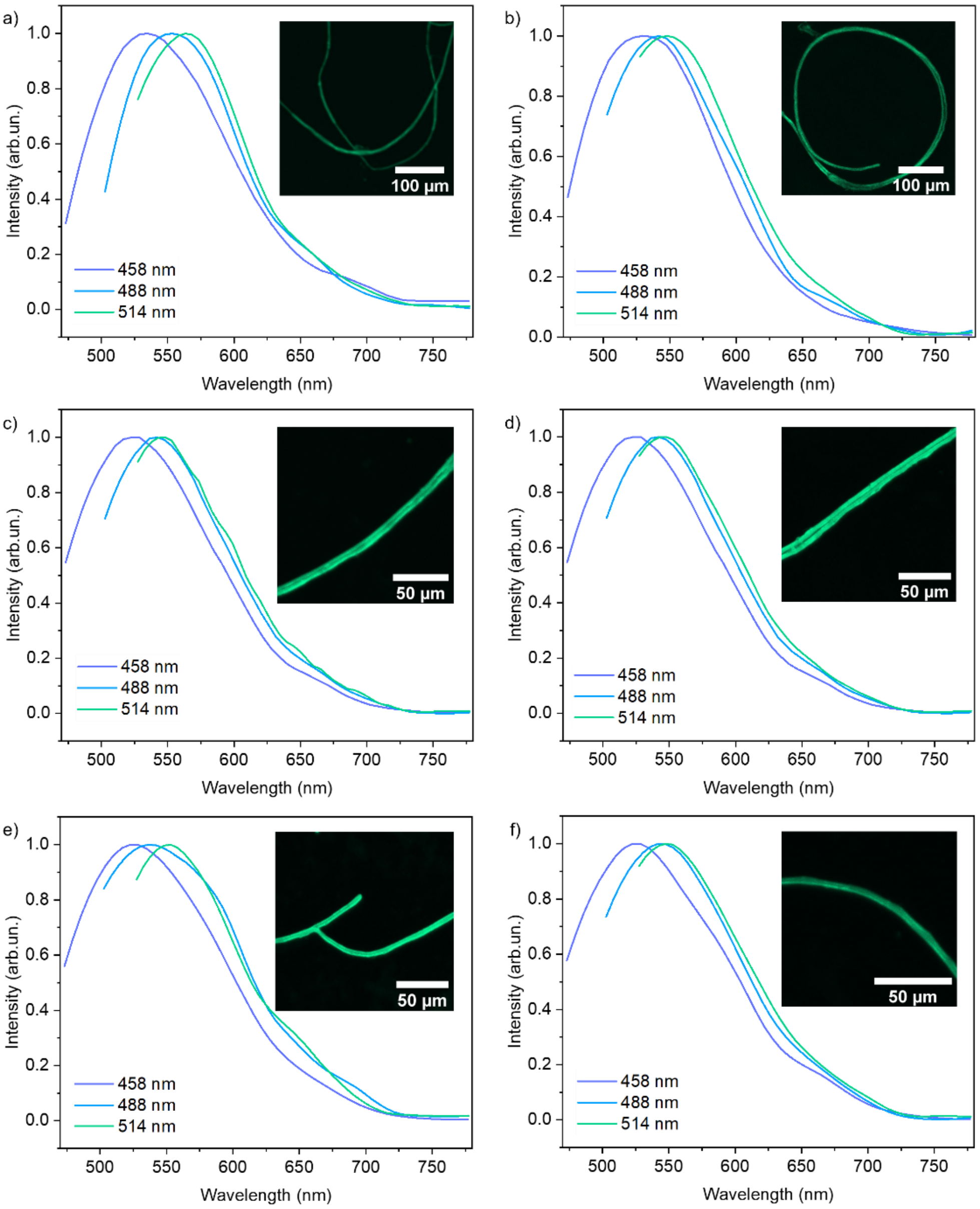

**Figure S5. Micro-photoluminescence properties of microfibers**. (a-f) Confocal fluorescence spectra of various microfibers as measured by optical excitation at 458, 488 and 514 nm. The fluorescence micrograph of each of the investigated microfibers is shown in the insets, respectively. The fluorescence micrographs were obtained by using a 375 nm laser for optical excitation.

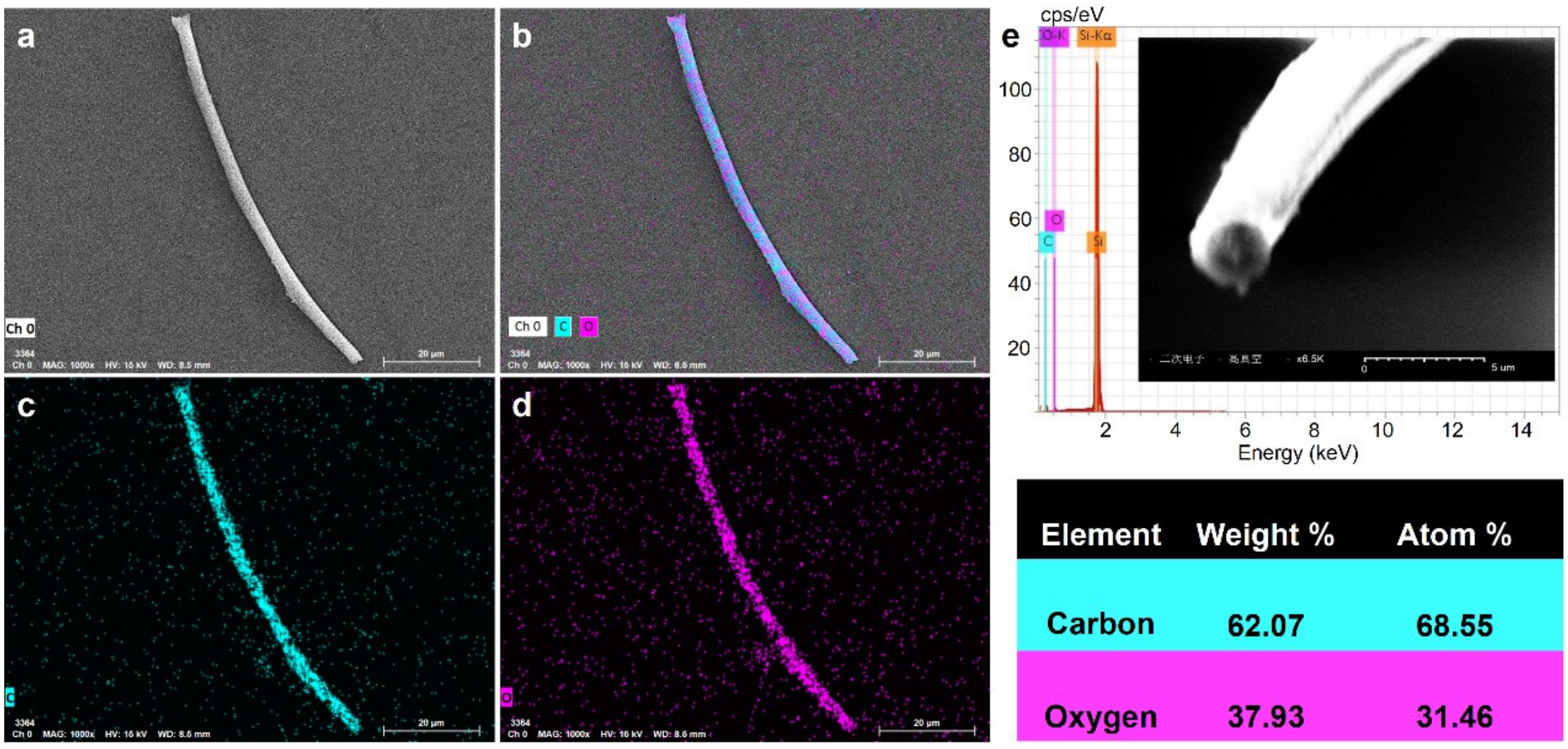

**Figure S6. SEM and EDS characterization of the single-filament microfiber.** (a) SEM micrograph of a microfiber. (b) EDS mapping overlapped with SEM morphology. (c, d) EDS mapping for element C and O, respectively. Scale bars: 20 μm. (e) EDS analysis of the microfiber with 68.55 % of Carbon and 31.46 % of Oxygen, respectively. Inset is the SEM image of the end facet for microfiber.

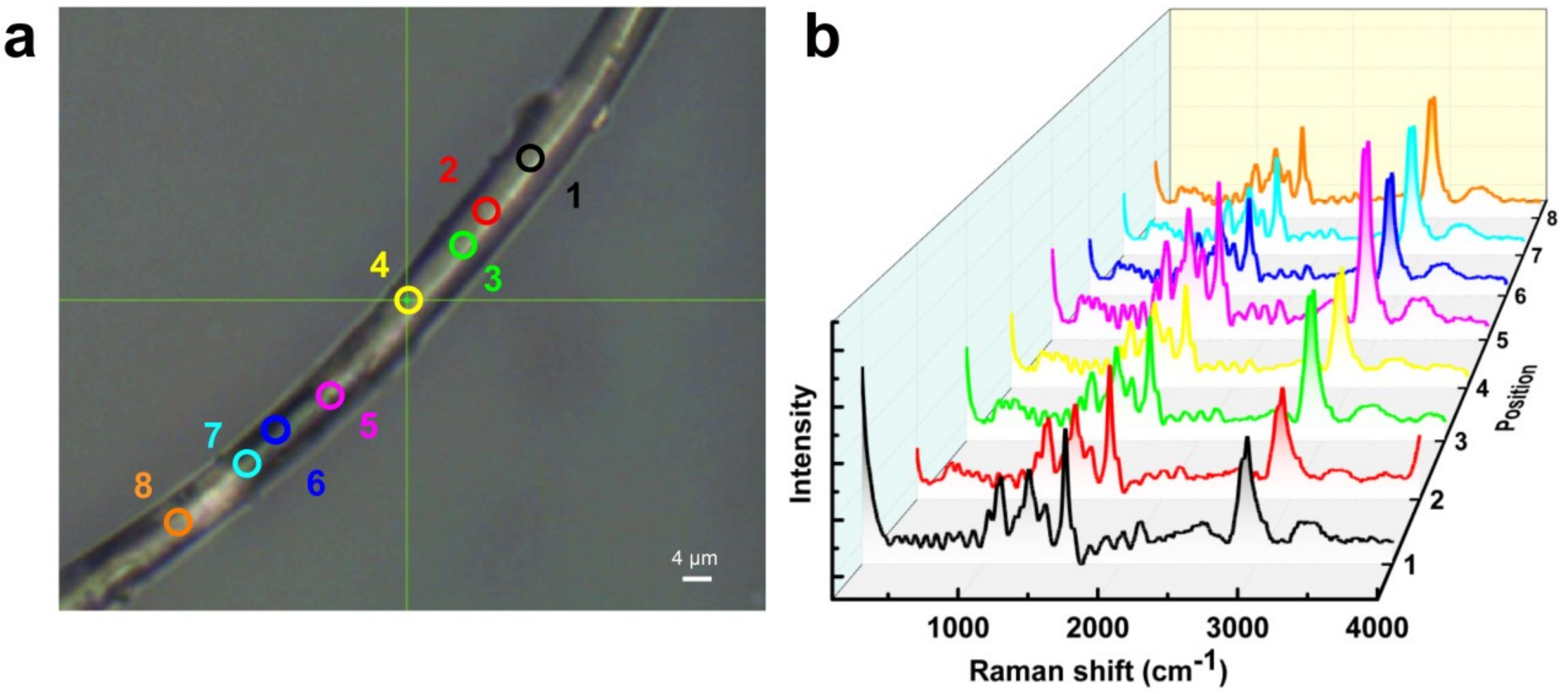

**Figure S7. Spatially-resolved micro-Raman characterization of the microfiber.** (a) Optical micrograph of a single microfiber. The acquisition points of micro-Raman spectra are highlighted by the colored circles. (b) The micro-Raman spectra corresponding to the acquisition points in (a).

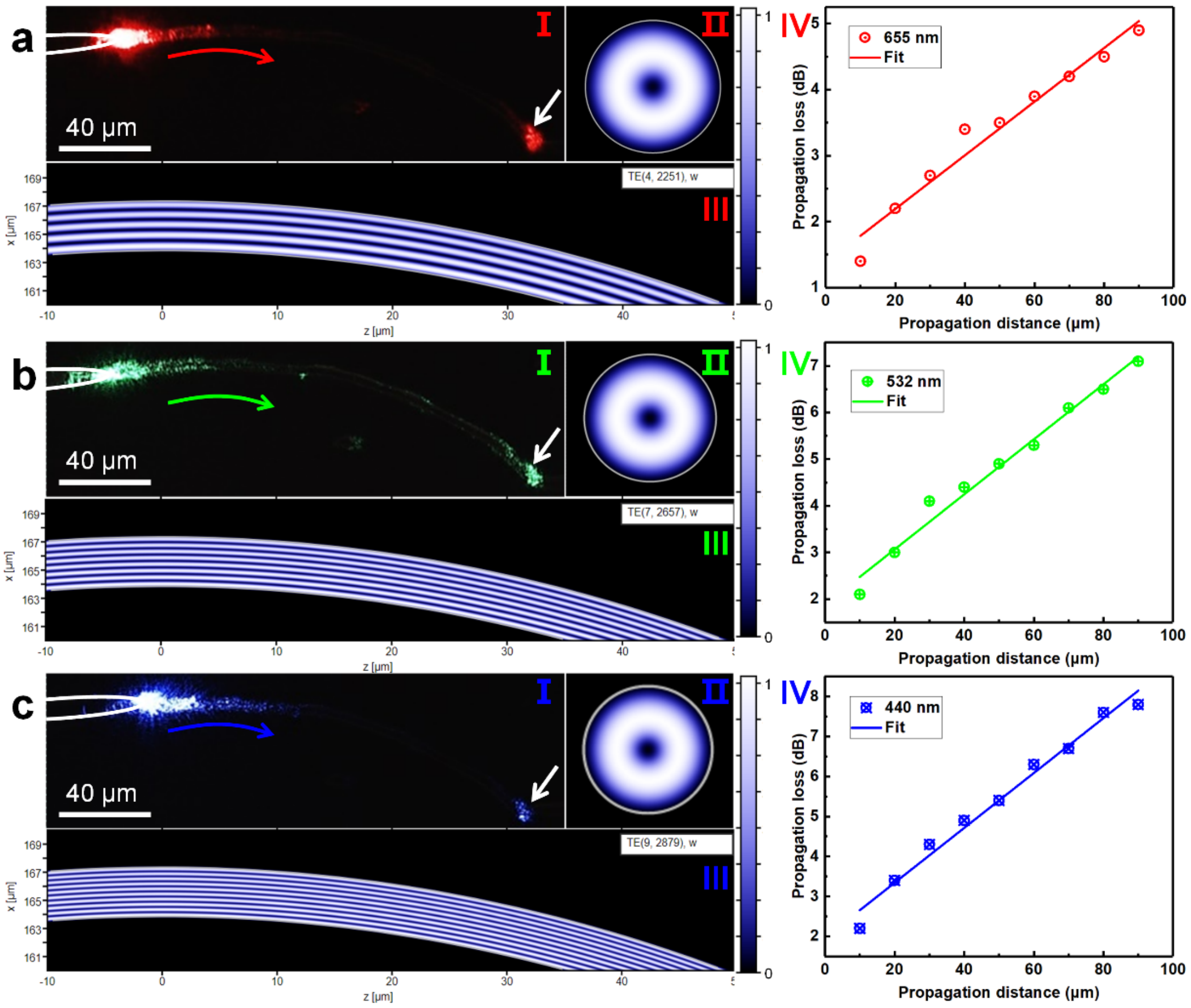

**Figure S8.** Simulated electromagnetic energy density (II cross section, III lateral section) and propagation loss (IV) distribution in the bended microfiber (I) guiding red (a), green (b) and blue (c) lasers. The calculated loss coefficients of 0.04, 0.06, 0.07 dB/μm are obtained for red, green, and blue lasers, respectively.

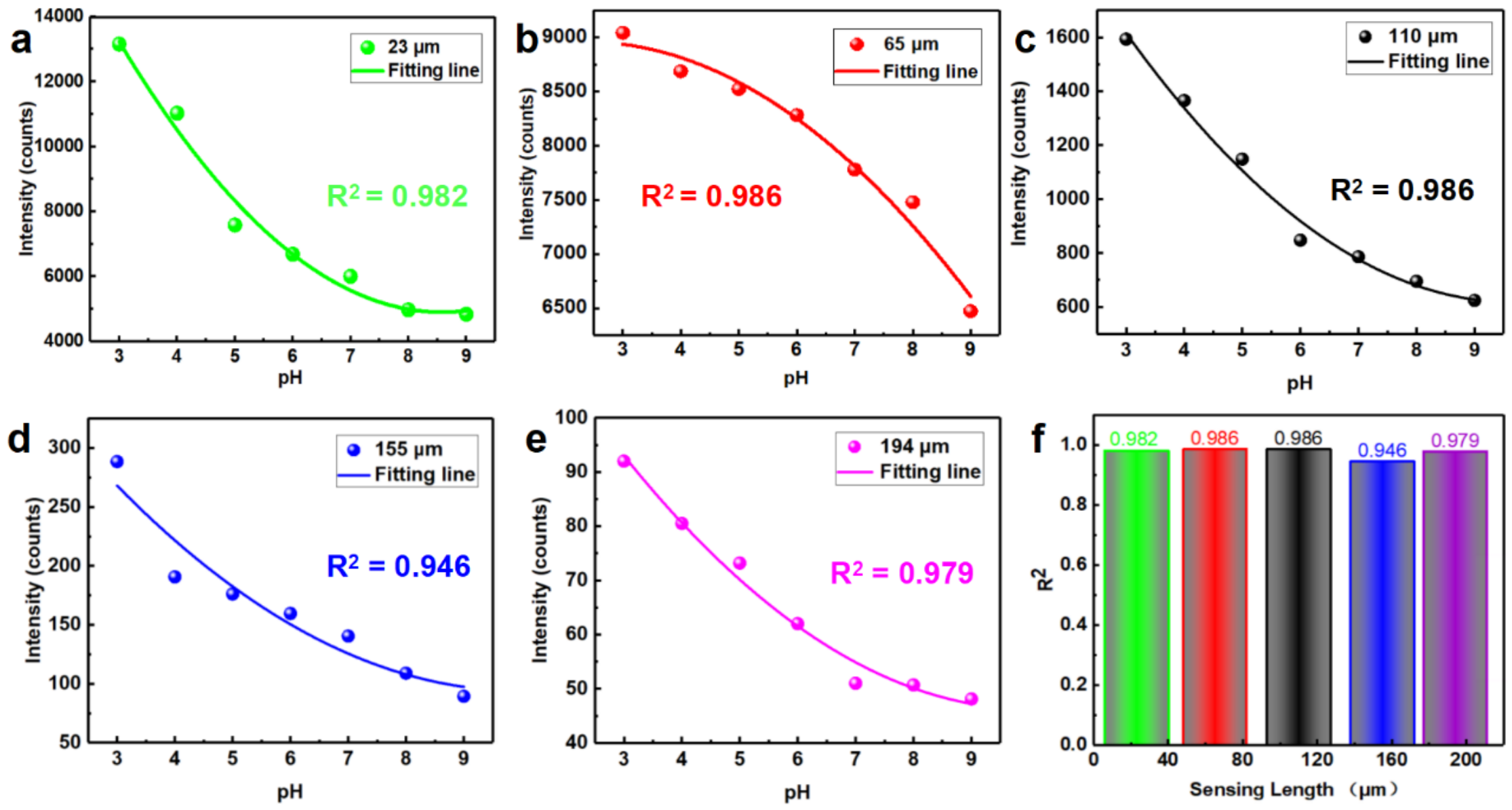

**Figure S9.** PL intensity of the microfiber in liquid microenvironments with different pH at different sensing lengths of (a) 23, (b) 65, (c) 110, (d) 155 and (e) 194 μm. Solid dots are the experimental data and solid line is the fitting curve. (f) The relationship between fit coefficient of $R^2$ and sensing length. $R^2$ is the fitted index. It indicates how well the model fits the data, with close to 0 indicating a worse fit and close to 1 indicating a better fit. Where the fitted equation is : $y = y0 + Ae^{R_0 * x}$ .

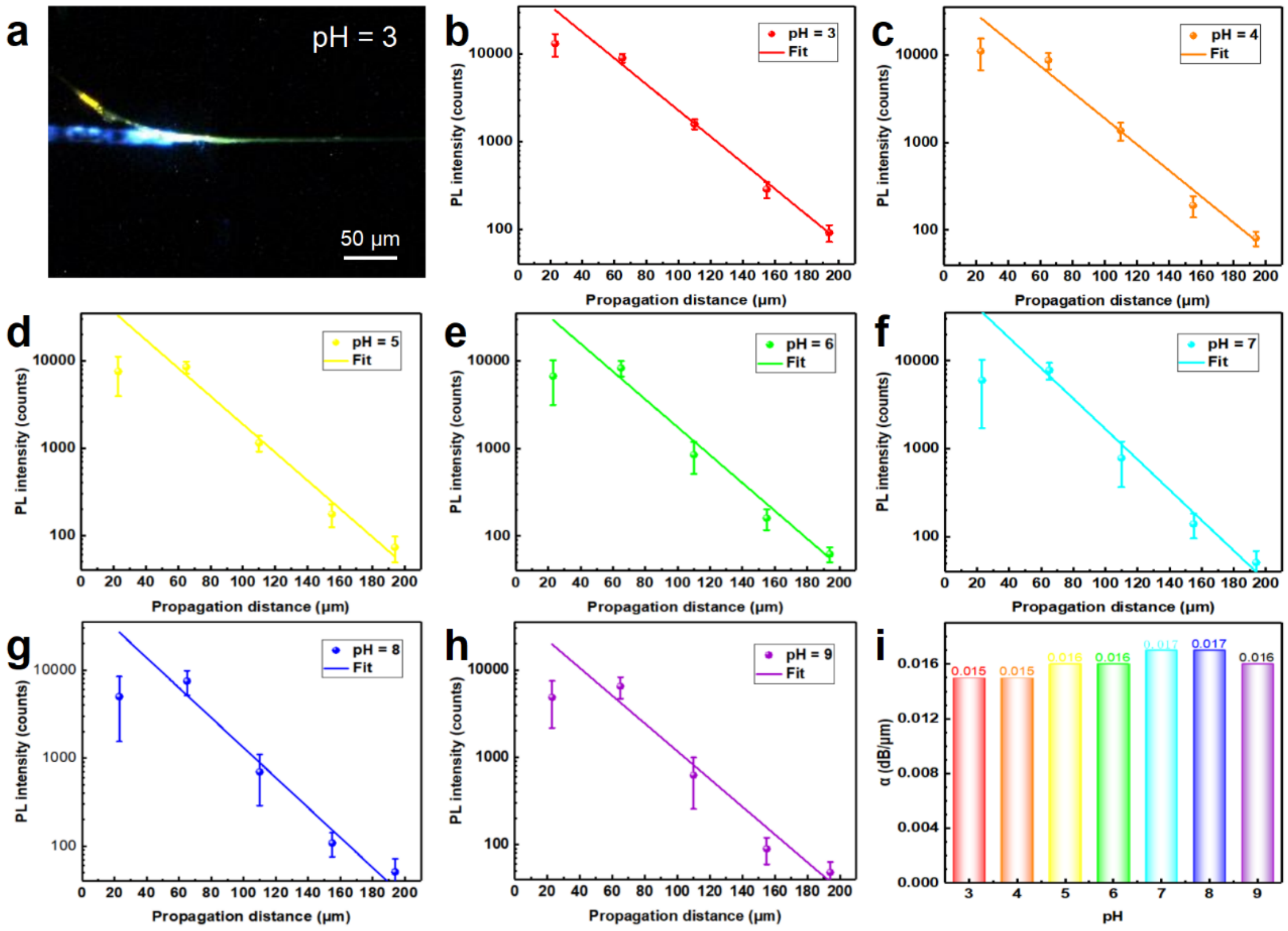

**Figure S10.** PL intensity of the microfiber at different propagation distance in liquid microenvironments with different pH of 3-9 (a, b-h). Solid dots are the experimental data and solid line is the exponential fitting curve. Error bar is the standard deviation for five independent measurements. (f) The relationship between loss coefficient, $\alpha$, and solution pH.

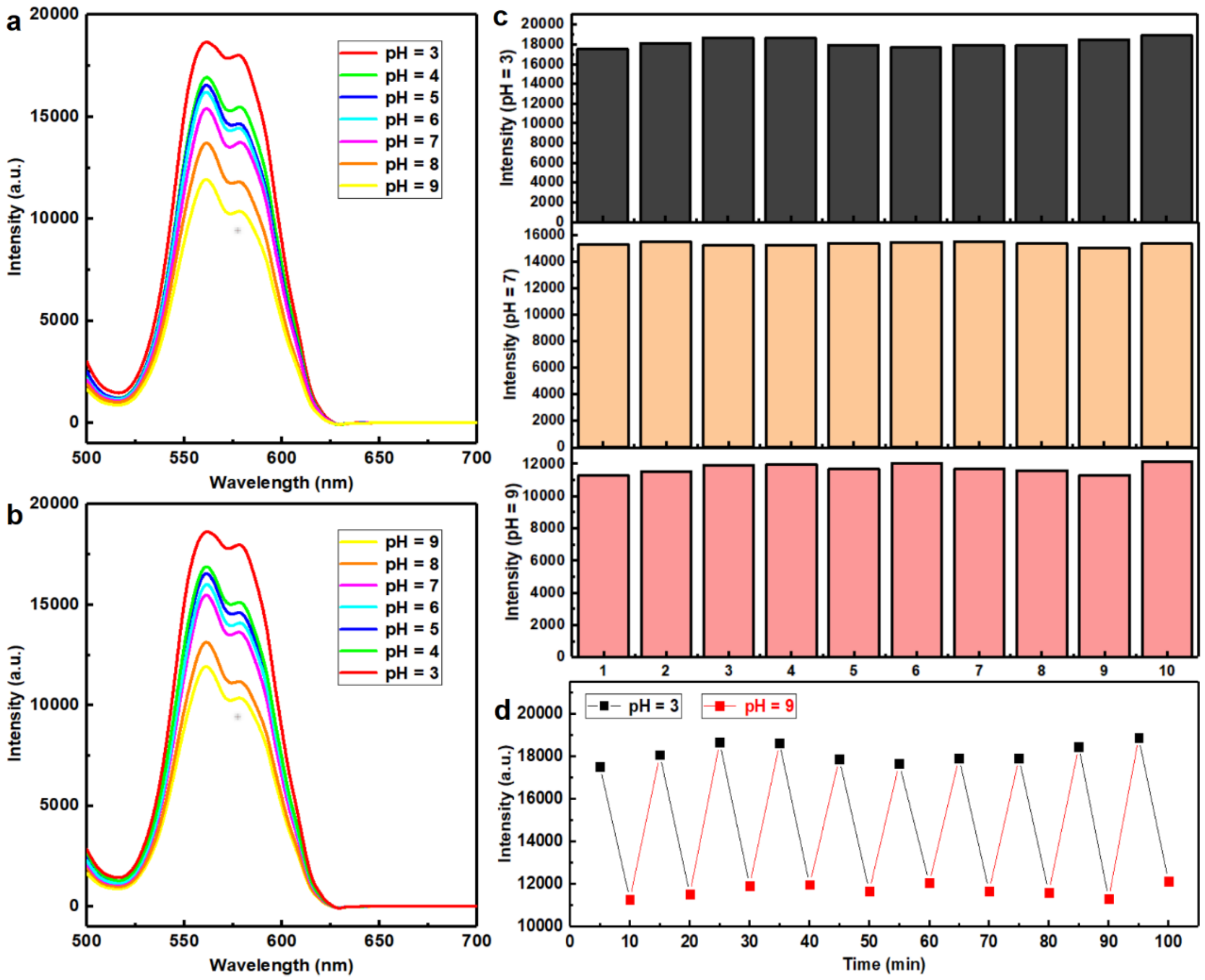

**Figure S11. The reversibility of the optical response of the single-filament microfiber.** (a) Photoluminescence (PL) spectra of a single-filament microfiber immersed in a solution with a pH varying from 3 to 9 upon blue light excitation with constant power. (b) PL spectra of the single-filament microfiber immersed in a solution with a pH varying from 9 to 3 upon blue light excitation at constant power. (c) PL intensity in ten consecutive cycles of pH solution variation from pH = 3 to pH = 9 and back to pH = 3. The graph shows the PL intensity of the single-filament microfiber for solutions with pH = 3, 7, and 9. (d) Changes in the PL intensity of the single-filament microfiber PL upon cyclic variation of pH solution between pH = 3 and pH = 9.

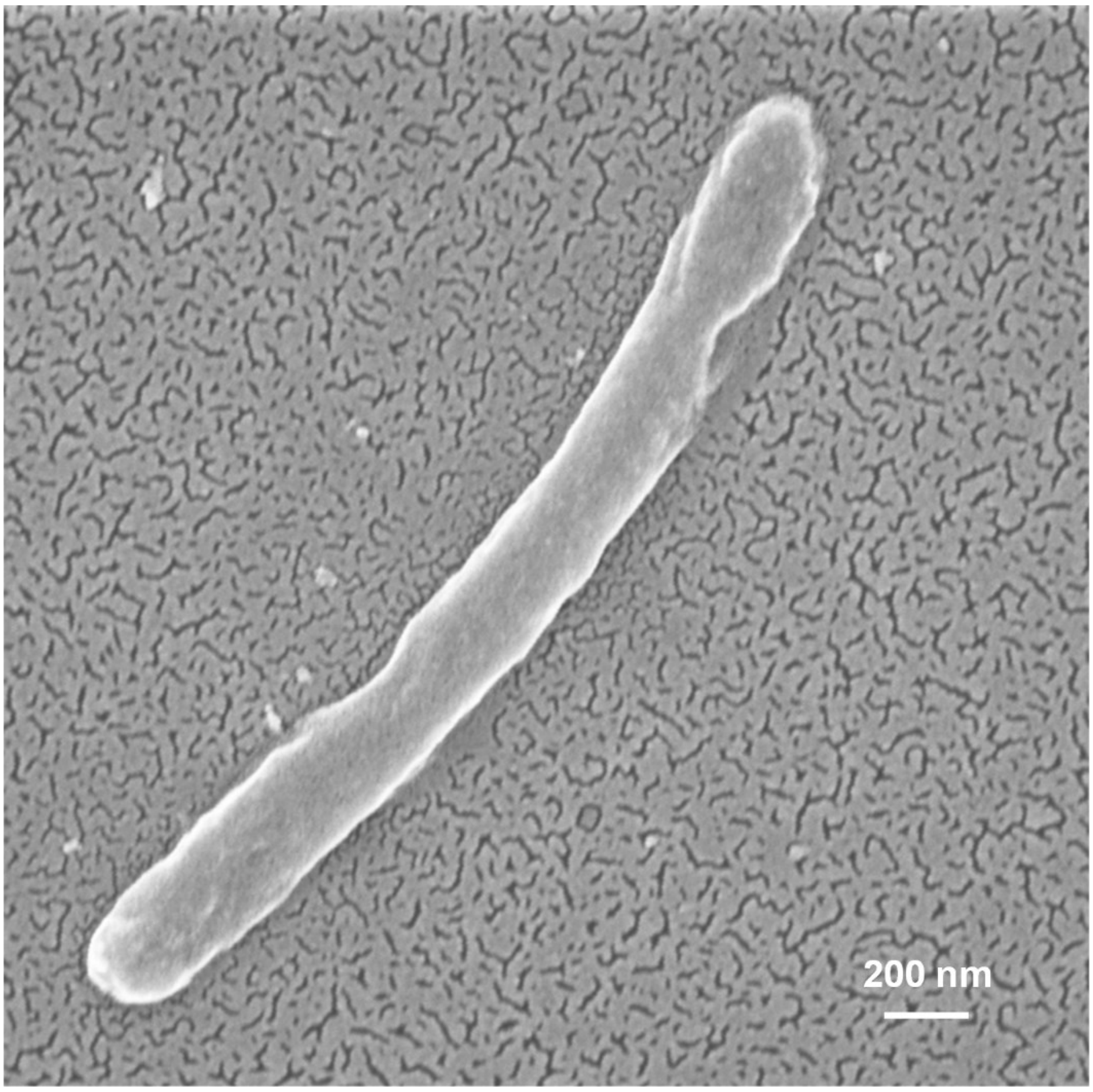

**Figure S12.** Scanning electron microscopy image of *Helicobacter pylori* used in our experiment.